# Anisotropic intermediate valence in $Yb_2M_3Ga_9$ (M = Rh, Ir)

A. D. Christianson[1,2], J. M. Lawrence[1], A. M. Lobos[3], A. A. Aligia[3], E. D. Bauer[2], N. O. Moreno[2], C. H. Booth[4], E. A. Goremychkin[5], J. L. Sarrao[2], J. D. Thompson[2], C. D. Batista[2], F. R. Trouw[2], M. P. Hehlen[2]

[1]University of California Irvine, California, 92697

[2]Los Alamos National Laboratory, Los Alamos, New Mexico, 87545

[3]Centro Atómico Bariloche and Instituto Balseiro, Comisión Nacional de Energía Atómica, 8400 S. C. de Bariloche, Argentina

[4]Lawrence Berkeley National Laboratory, Berkeley California, 94550

[5]Argonne National Laboratory, Argonne Illinois, 60439

The intermediate valence compounds $Yb_2M_3Ga_9$ (M = Rh, Ir) exhibit an anisotropic magnetic susceptibility. We report measurements of the temperature dependence of the 4f occupation number, $n_f(T)$, for $Yb_2M_3Ga_9$ as well as the magnetic inelastic neutron scattering spectrum $S_{mag}(\Delta E)$ at 12 and 300 K for $Yb_2Rh_3Ga_9$. Both $n_f(T)$ and $S_{mag}(\Delta E)$ were calculated for the Anderson impurity model with crystal field terms within an approach based on the non-crossing approximation. These results corroborate the importance of crystal field effects in these materials; they also suggest that Anderson lattice effects are important to the physics of $Yb_2M_3Ga_9$.

Most of the metallic intermediate valence (IV) compounds such as $CeSn_3$, $CePd_3$, and $YbAl_3$ are cubic and exhibit isotropic behavior. For these materials where the Kondo temperature is large ($T_K \geq 500K$) it is not necessary to invoke either magnetic RKKY interactions or crystal field (CF) effects to explain the physical behavior. Recently, Moreno *et al.*[1] have shown that CF interactions play an important role in the physics of the IV compounds $Yb_2M_3Ga_9$. This is manifested by a difference in magnitudes, Weiss constants and temperatures of the maximum for the in-plane and out-of-plane magnetic susceptibility. The anisotropic susceptibility was reproduced by calculations using the Zwicknagl, Zevin and Fulde (ZZF) simplification[2] of the non-crossing approximation (NCA) to the Anderson impurity model (AIM), including the effects of CF interactions. (We note that the ZZF method for an AIM including crystal fields has been applied previously to the nearly-trivalent compound $YbN^3$).

In this paper we report results of measurements of the temperature dependence of the 4f occupation number $n_f(T)$ for $Yb_2M_3Ga_9$ measured using $L_{III}$-edge x-ray absorption[4] and also the results of inelastic neutron scattering (INS) measurements[5] on $Yb_2Rh_3Ga_9$. We compare the results of these measurements and of the earlier[1] measurements of the susceptibility $\chi(T)$ to the results of calculations based on the ZZF approach.

$Yb_2M_3Ga_9$ compounds crystallize in a hexagonal structure (space group $P6_3cm$).[6] For the INS measurements, approximately 40 g of polycrystalline $Yb_2Rh_3Ga_9$ and $Y_2Rh_3Ga_9$ were prepared by the following procedure. Stoichiometric amounts of Yb, Rh, and Ga were placed in an alumina crucible and sealed in a silica tube under vacuum. In the case of $Y_2Rh_3Ga_9$, stoichiometric amounts of Y and Rh were first arc-melted. The samples were heated to 1150 °C, kept there for 1 hour, cooled to 1100 °C at a rate of 50 °C/hr, then cooled to 900 °C at a rate of 150 °C/hr, kept at 900 °C for 1-2 days, at which point the samples were quenched in liquid nitrogen. X-ray diffraction confirmed the hexagonal structure (only small impurity peaks were observed corresponding to less than 5% of the sample volume) and magnetic susceptibility measurements were consistent with single crystal results. The resulting samples were then ground into powder and placed in a flat plate sample holder. Two time-of-flight chopper spectrometers were used: LRMECS at the Intense Pulsed Neutron Source at Argonne National Laboratory and PHAROS at the Los Alamos Neutron Science Center at Los Alamos National Laboratory. The experimental configuration for these experiments was similar to that previously reported.[7] [8] Incident energies, $E_i$, of 15, 35, and 120 meV were used on LRMECS at 12 and 300 K and 60 meV was used at 12 K. On PHAROS $E_i$ of 70 and 120 meV were used at 18 and 300 K.

The magnetic contribution to the INS spectrum ($S_{mag}$) was determined using a standard method,[5 7] which assumes that the nonmagnetic scattering, (which dominates the scattering at large momentum transfer Q or high scattering angle) can be scaled to low angle (where the magnetic scattering dominates) in the same manner as observed experimentally in the nonmagnetic analogue compound $Y_2Rh_3Ga_9$.

For the $L_{III}$ edge x-ray absorption measurements, single crystals were grown as in Ref. 1. These measurements were performed at beamline 4-3 of the Stanford Synchrotron Radiation Laboratory and were analyzed using standard procedures[4] to obtain $n_f(T)$.

Fig. 1 displays minimally processed data (corrected only for counting time) collected for $Yb_2Rh_3Ga_9$ and $Y_2Rh_3Ga_9$. In (a) the temperature dependence of the low Q INS data collected on PHAROS with $E_i = 120$ meV for $Yb_2Rh_3Ga_9$ is shown. A noticeable decrease in the scattering for $40 < \Delta E < 90$ meV is observed at 300 K relative to 18 K. (The small feature at ~90 meV is an artifact of the empty holder scattering which has not been subtracted out.) In (b) the low Q INS data with $E_i = 120$ meV for $Yb_2Rh_3Ga_9$ collected on LRMECS (corrected for sample mass, nuclear scattering cross-section, neutron absorption, and counting time) is compared to that for $Y_2Rh_3Ga_9$ at 12 and 300 K. Magnetic scattering is evidenced by the stronger scattering in the Yb compound than in the Y compound. The scattering for $Y_2Rh_3Ga_9$ at 300 K is greater than for 12 K when $\Delta E < 30$ meV. For $\Delta E > 30$ meV the scattering is comparable at both 300 and 12 K. This type of temperature dependence is as expected for phonon scattering. In contrast, in $Yb_2Rh_3Ga_9$ the LRMECS and the PHAROS data reveal the scattering at 300 K is actually *less* than at 12 K for $40 < \Delta E < 80$ meV. Hence, the temperature dependence can *not* be explained solely by the temperature dependence of phonon scattering but is as expected for CF excitations, where the occupation of the ground state multiplet decreases with temperature, and/or for the transfer of spectral weight from an inelastic response at low temperature to a quasielastic response at high temperatures.

Fig. 2 displays $S_{mag}$ at 12 and 300 K for data collected on LRMECS for a variety of $E_i$'s (in addition to the corrections listed for fig. 1(b) the contribution of the empty sample holder has been removed and the nonmagnetic background as described earlier). The Lorentzian fits in Fig. 2(a) and (b) show that the INS spectrum does indeed evolve from a high temperature quasielastic response to an inelastic response at low temperature. This is characteristic of IV systems.[5] At 12 K (Fig. 2(a)) the data can be fit with a single Lorentzian with width 27(1) meV and position 27(1) meV. At 300 K (Fig. 2(b)) the data can be fit with a single Lorentzian centered at $\Delta E = 0$ meV (quasielastic) with width 20.7(4) meV. Values of the static susceptibility can then be derived from these fits at 12 and 300 K and compared to the polycrystalline average of the measured susceptibility (Fig. 2(c)). At 300 K the comparison is quite reasonable, but at 12 K there is a discrepancy between the value of the static susceptibility derived from the INS results and those determined by the bulk measurements. Evidently, the INS measurements reported here are not sensitive to the low temperature upturn in the susceptibility— neglecting this upturn results in more reasonable agreement. This suggests that the upturn is an extrinsic effect unless its contribution is only at low energies and thus not observable in these investigations.

The experimental $n_f(T)$ results from 20 to 550 K are displayed in Fig. 3. The temperature dependence and magnitude of $n_f$ is consistent with other Yb IV systems. The data indicate a somewhat higher degree of fractional valence in $Yb_2Ir_3Ga_9$ than in $Yb_2Rh_3Ga_9$.

We now turn to a discussion of systematic errors in our determination of $S_{mag}$. We have used a standard procedure[5 7] to determine the nonmagnetic scattering. Specifically, we have used the nonmagnetic analog $Y_2Rh_3Ga_9$ to determine the scaling factor between the high and low angle data and then have applied that same scaling factor to the $Yb_2Rh_3Ga_9$ data. Systematic error could be introduced should the scaling factor depend significantly on the nuclear scattering cross section, which differs for Y and Yb. Since the magnetic susceptibility derived from the INS analysis compares favorably with that determined from bulk measurements, we argue that this has only a small (~10%) effect on our results. A small systematic error also could be introduced due to the strong neutron absorption in this material. We have minimized this by using a flat plate sample geometry, and correcting for the effects of neutron absorption.

Systematic error in the value of the $n_f(T)$ determined by $L_{III}$ measurements is discussed in some detail in Ref. 4, where it is estimated to have an effect on the absolute value of $n_f$ that is no larger than 5-10%. Such error should be independent of temperature and hence should not affect the temperature dependence determined for $n_f$.

The AIM calculations that have been performed for the various experimental quantities shed additional light on the physics in the $Yb_2M_3Ga_9$ materials. These calculations are an extension of those reported in Ref. 1; for the INS calculations we used Eq. 39 of Zwicknagl et al.[2] with the same spectral density used in Ref. 1. In the lowest order calculation (parameter set 1, Table I) the CF splits the 8-fold degenerate ground state of $Yb^{3+}$ into two quartets separated by the energy $\Delta$. The CF wave functions are taken as eigenstates of the angular momentum operator $J_z$, where the ground state quartet consists of two doublets, $|\pm 7/2\rangle$ and $|\pm 5/2\rangle$, and the excited state quartet consists of the other two doublets, $|\pm 3/2\rangle$ and $|\pm 1/2\rangle$. This scheme gives a larger susceptibility for H//c than for H//a, due to the larger angular momentum (along the *c*-axis) in the ground state multiplet. By choosing the hybridization between the 4f and conduction electrons to have a value $\Gamma$ = 165 K and the CF splitting to have a value $\Delta$ = 400 K, we reproduce the temperature of (~100 K) of the peak in $\chi_c(T)$, which arises from the Kondo effect ($T_K$ = 194 K) in a level of effective degeneracy four at low temperatures, and the temperature (~150 K) of the peak in $\chi_{ab}(T)$, which arises from the CF excitations. The fit to the susceptibility and $n_f(T)$ (dashed lines in Fig. 2(c) and lines in Fig. 3) is qualitatively quite good. With the inclusion of a mean field interaction parameter I ($1/\chi_{tot} = 1/\chi_{NCA} + I$) this scheme gives excellent quantitative fits to the susceptibility (solid lines, Fig. 2(c)).

When calculating the INS spectrum the same mean field interaction parameter as used for the static susceptibility was included, as a reduction factor independent of energy. The fit to the INS (Fig. 2(a), dash-dot line) is of the right magnitude and overall energy scale, but shows two features, one at 20 meV representing the Kondo scattering in the ground state quartet and the other at 55 meV representing the CF excitation which due to the Kondo effect occurs at $k_B(\Delta + T_K)$. The fits to the INS can be improved (Fig. 2(a), dashed line) without affecting the susceptibility or occupation number by lifting the ground state degeneracy to two doublets separated by 100 K, lifting the degeneracy of the upper quartet to two doublets separated from the ground state by 280 and 340 K, and allowing the hybridization constant $\Gamma$ to be slightly different for the ground and excited multiplets (parameter set 2, Table I).   Overall, the combination of CF and Kondo physics gives good semi-quantitative fits to the anisotropic susceptibility, $n_f(T)$ and INS spectrum.  In this sense, our calculations capture much of the essential physics of these compounds.

Although at zero temperature the ZZF approximation[2] is equivalent to a variational approximation that is exact for infinite degeneracy, the temperature dependence of the susceptibility is governed by only one energy scale ($T_K$), whereas in IV systems it should also depend on the charge fluctuation energy $E_c$.  Given the strong mixed valence of these compounds ($n_f(0) \sim 0.6$), one expects that $E_c$  (which is of the order of 2000 K in both compounds) should play a role in the temperature dependence.  However, as shown by Bickers, Cox and Wilkins,[9] the susceptibility for $n_f = 0.7$ differs from the one-parameter Kondo function by no more than 5%.   We ignore this difference because it is smaller than the amount that the susceptibility is reduced by inclusion of the molecular field parameter I = 20 mol/emu (20% for $\chi_c$ and 10% for $\chi_{ab}$).  Such a parameter can arise due to antiferromagnetic interactions, but the Néel temperature $T_N$ = 52 K deduced from the formula I = $T_N$/ $C_{7/2}$ (where $C_{7/2}$  = 2.58 emu-K/mol is the Yb Curie constant) is much larger than the ordering temperatures (~1K) seen in magnetic Yb compounds. One possibility is that the large value of I needed to improve the agreement between the magnitude of the measured susceptibility and the value calculated in the AIM is not due to normal RKKY interactions but is an Anderson *lattice* effect, where antiferromagnetic interactions are enhanced due to the mixed valence.

The existence of Anderson lattice effects is also suggested by the fact that our calculation significantly overestimates $n_f(300 \text{ K})$ (Fig. 3). Similar behavior of $n_f(T)$ in cubic IV compounds has been interpreted as evidence that the crossover from low temperature Fermi liquid behavior to high temperature local moment is *slower* in the Anderson lattice than in the AIM.[10] This effect arises primarily from the valence change that occurs as a function of temperature: for the Yb $4f^{14-nf}(5d6s)^{2+nf}$ configuration, $z = 2 + n_f(T)$ and Anderson/Kondo physics causes $n_f(T) \rightarrow 1$ for $T > T_K$ ($n_f(T) \rightarrow (2J+1)/(2J+2)=8/9$ for $T > E_c$). For the Anderson lattice, a finite concentration of extra valence electrons $\Delta z = z(\infty) - z(0)$ must be accommodated by the conduction band. The resulting increase in kinetic energy is reduced by reducing the rate of the valence change with temperature, giving a "slow crossover." This is not the case for the AIM, where only one degree of freedom is involved.

The calculation of the INS at 300 K using parameter set 2 (the largest magnitude dashed curve in fig. 2(b)) also overestimates the data. We have calculated $S_{mag}$ at 300 K using smaller values of $n_f(0)$ in parameter set 2 (Fig. 2(b)). By choosing the smaller $T = 0$ value $n_f(0) = 0.5$ that yields, in the context of the AIM, the actual experimental value of $n_f(300 \text{ K}) = 0.80$, we obtain much better agreement with $S_{mag}$ at 300 K which suggests that the slow crossover affects the temperature evolution of $S_{mag}$.

We have measured $n_f(T)$ for $Yb_2Rh_3Ga_9$ and $Yb_2Ir_3Ga_9$ as well as the INS response of $Yb_2Rh_3Ga_9$ at 12 and 300 K. The experimental results for $n_f(T)$ definitively establish that $Yb_2M_3Ga_9$ are IV systems. The INS spectra are similar to those seen in other IV systems, where the spectral weight shifts from a broad quasielastic response at high temperatures to a broadened inelastic response at low temperatures. We have performed AIM calculations which corroborate the results of Ref. 1 that crystal fields play an important role in the physics of $Yb_2M_3Ga_9$ and that the AIM, with CF splitting, captures much of the essential physics. In addition, our results suggest that Anderson lattice effects must be considered to explain the temperature dependence of $n_f(T)$ and of the INS spectra; in particular, the temperature dependence suggests the presence of a slow crossover to the local moment regime.

Work at Irvine was supported by the Department of Energy (DOE) under Grant No. DE-FG03-03ER46036. Work at Los Alamos, Argonne, and Lawrence Berkeley (Contract No. DE-AC03-76SF00098) was performed under the auspices of the DOE. Portions of this research were carried out at the Stanford Synchrotron Radiation Laboratory. This work was sponsored by PICT 03-12742 of ANPCyT. A. L. and A. A. A. are partially supported by CONICET.

## Tables

| | E(\|±7/2⟩) | E(\|±5/2⟩) | E(\|±3/2⟩ | \|1/2⟩) | $\Gamma_{\pm7/2,\pm5/2}$ | $\Gamma_{\pm3/2,\pm1/2}$ | I | $T_K$ | $\gamma$ | $n_f(0)$ |
|---|---|---|---|---|---|---|---|---|---|---|
| $Yb_2Rh_3Ga_9$ 1 | 0 | 0 | 400 | 400 | 165 | 165 | 0 | 194 | 68 (45) | 0.59 |
| $Yb_2Rh_3Ga_9$ 2 | 0 | 100 | 280 | 340 | 180 | 190 | 20 | 187 | 61 (45) | 0.60 |
| $Yb_2Ir_3Ga_9$ | 0 | 0 | 280 | 280 | 230 | 230 | 45 | 1030 | 28 (25) | 0.53 |

**Table I:  Fit parameters for AIM calculations for $Yb_2M_3Ga_9$.  For $Yb_2Rh_3Ga_9$, 1 and 2 refer to parameter sets 1 and 2.  The parameters for $Yb_2Ir_3Ga_9$ are from ref. 1. The first four columns correspond to the energy of the CF levels, $\Gamma_{\pm7/2,\pm5/2}$ and $\Gamma_{\pm3/2,\pm1/2}$ denote the conduction electron hybridization to the specified CF levels, I is a mean field interaction parameter, $T_K$ is the Kondo temperature,  $\gamma$ is the Sommerfeld parameter where the values in parentheses are experimental values from ref. 1, and $n_f(0)$ is the 4f occupation at T = 0 K.  All units are in K except for I ((mol-Yb)/emu), $\gamma$ (mJ/(mol Yb $K^2$)), and $n_f(0)$ (dimensionless).**

**Figure Captions**

**Fig. 1. (a) Comparison of data collected with PHAROS for $Yb_2Rh_3Ga_9$ at 18 (circles) and 300 K (squares) with $E_i = 120$ meV.  (b)  Comparison of data collected with LRMECS for $Yb_2Rh_3Ga_9$ and $Y_2Rh_3Ga_9$ with $E_i = 120$ meV.  The squares (circles) denote the scattering for $Yb_2Rh_3Ga_9$ ($Y_2Rh_3Ga_9$) at 12 K.  The solid (dashed) line denote the scattering for $Yb_2Rh_3Ga_9$ ($Y_2Rh_3Ga_9$) at 300 K.  In both (a) and (b) the statistical errors bars are approximately the symbol size.**

**Fig. 2. Comparison of $S_{mag}$ at 12 and 300 K and the magnetic susceptibility.  (a) displays $S_{mag}$ at 12 K with $E_i = 15$ (stars), 35 (circles), 60 (squares), and 120 meV (triangles).  The solid line is a Lorentzian fit and the dash-dot (dashed) line is an AIM calculation for parameter set 1 (2) as described in the text.  (b) displays $S_{mag}$ at 300 K with $E_i = 15$ (stars), 35 (circles), and 120 (triangles) meV.  The solid (dashed) lines are a Lorentzian fit (AIM calculations) with parameters as described in the text.  The upper most dashed curve corresponds to $n_f(0) = 0.6$, the next lower $n_f(0) = 0.5$, and the lowest $n_f(0) = 0.4$. (c) displays the magnetic susceptibility for field applied along the c-axis (circles), field applied in the basal plane (triangles), and the polycrystalline average (squares).  The stars denote the values of the static susceptibility derived from Lorentzian fits to $S_{mag}$ at 12 and 300 K.  The dashed (solid) lines denote AIM calculations of the static susceptibility for parameter sets 1 (2).**

**Fig. 3.** $n_f(T)$ for $Yb_2Rh_3Ga_9$ (squares) and $Yb_2Ir_3Ga_9$ (triangles). Lines are AIM calculations as described in the text.

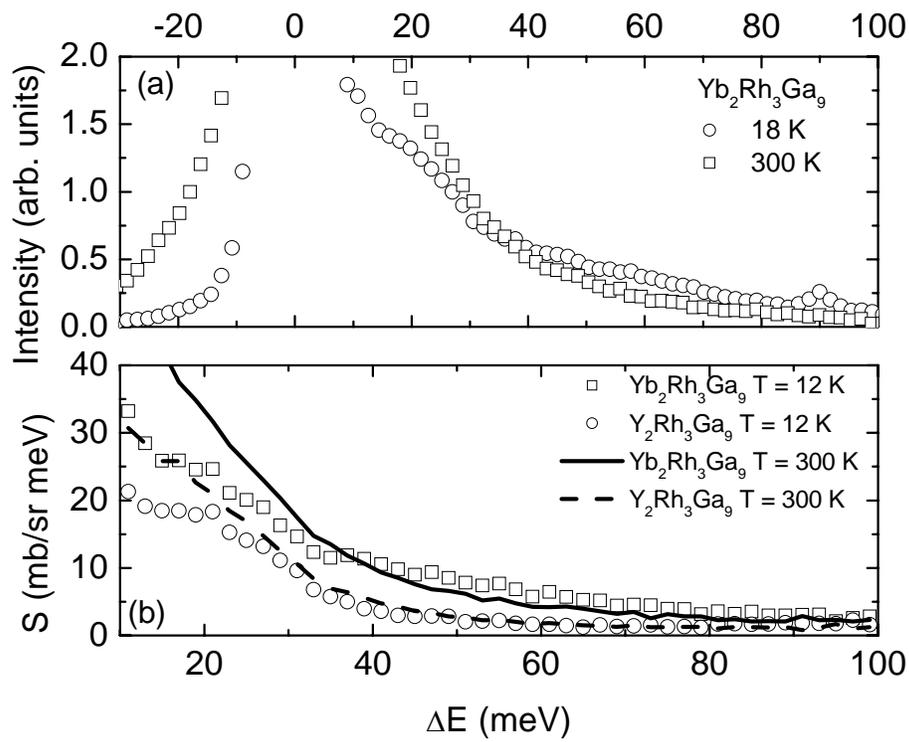

Figure 1.

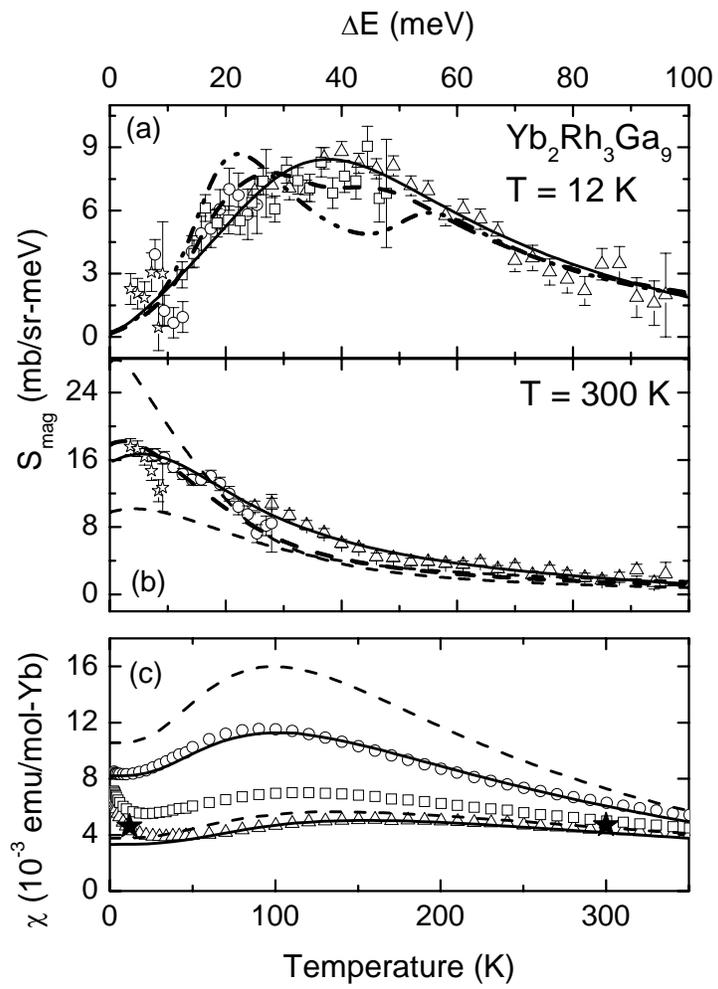

Figure 2.

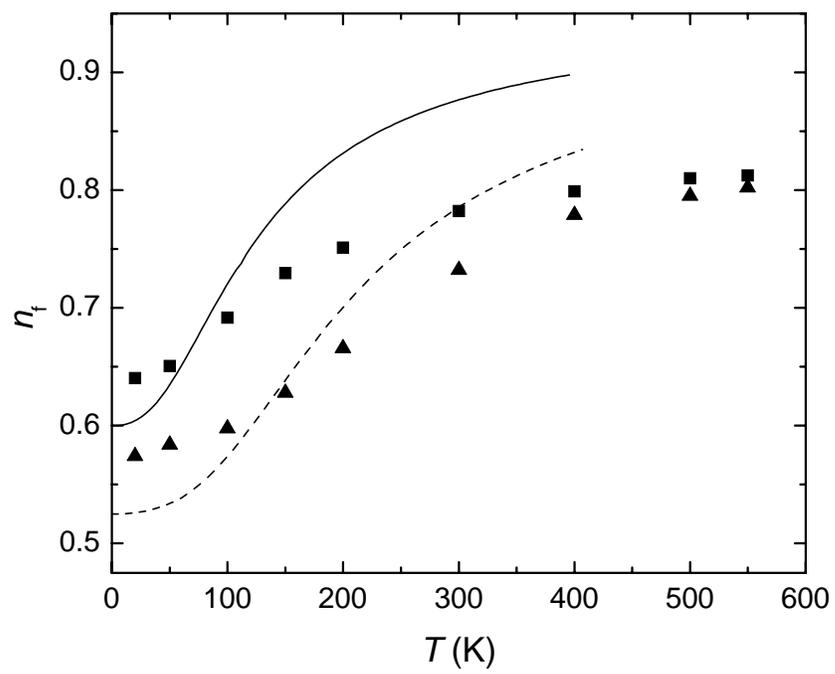

Figure 3.